\documentclass[a4paper,11pt]{article}

\usepackage{graphicx}
\usepackage{xcolor}
\usepackage{todonotes}
\usepackage{authblk}
\usepackage{amsmath}
\usepackage{amssymb}
\usepackage{amsfonts}

\usepackage{subcaption}
\usepackage{url}

\providecommand{\keywords}[1]{\textit{Keywords:} #1}

\begin{document}

\title{An efficient geometric method for incompressible hydrodynamics on the sphere}

\author[1,2]{Paolo Cifani}
\author[3]{Milo Viviani}
\author[4]{Klas Modin}

 \affil[1]{Multiscale Modeling and Simulation, Faculty EEMCS, University of Twente, P.O. Box 217, 7500 AE Enschede, The Netherlands}
 \affil[2]{Gran Sasso Science Institute, Viale F. Crispi, 7 67100 L’Aquila, Italy}
 \affil[3]{Scuola Normale Superiore di Pisa, Pisa.}%
\affil[4]{Department of Mathematical Sciences, Chalmers University of Technology and University of Gothenburg, 412 96 Gothenburg, Sweden.}%

\date{\today}

\maketitle

\begin{abstract}
    We present an efficient and highly scalable geometric numerical method for two-dimensional ideal fluid dynamics on the sphere. The starting point is Zeitlin's finite-dimensional model of hydrodynamics. The efficiency stems from exploiting a tridiagonal splitting of the discrete spherical Laplacian combined with highly optimized, scalable numerical algorithms. For time-stepping, we adopt a recently developed isospectral integrator able to preserve the geometric structure of Euler's equations, in particular conservation of the Casimir functions. To overcome previous computational bottlenecks, we formulate the matrix Lie algebra basis through a sequence of tridiagonal eigenvalue problems, efficiently solved by well-established linear algebra libraries. The same tridiagonal splitting allows for computation of the stream matrix, involving the inverse Laplacian, for which we design an efficient parallel implementation on distributed memory systems. The resulting overall computational complexity is $\mathcal{O}(N^3)$ per time-step for $N^2$ spatial degrees of freedom. The dominating computational cost is matrix-matrix multiplication, carried out via the parallel library ScaLAPACK. Scaling tests show approximately linear scaling up to around $2500$ cores for the matrix size $N=4096$ with a computational time per time-step of about $0.55$ seconds. These results allow for long-time simulations and the gathering of statistical quantities while simultaneously conserving the Casimir functions. We illustrate the developed algorithm for Euler's equations at the resolution $N=2048$.
\end{abstract}


\section{Introduction}

Two-dimensional ideal hydrodynamics possess geometric properties that affect its qualitative long-time behaviour. 
In his seminal work, Arnold \cite{Arnold1966} found that ideal fluids can be described as a Hamiltonian system on the cotangent bundle of the infinite-dimensional Lie group of volume-preserving diffeomorphisms.
For two-dimensional fluids, this observation implies the existence of infinitely many first integrals, called \emph{Casimir functions} \cite{Marsden_book,Holm_book}.
Physically, they state the conservation of the integrated powers of vorticity and reflect that vorticity is advected along stream-lines. 
There is strong evidence that the long-time qualitative dynamics of non-viscous two-dimensional fluids is tied to the conservation of Casimirs, both enstrophy and higher order Casimirs~\cite{AbMa2003, Cifani2022}.
To better capture the correct long-time behaviour it is therefore natural to  look for numerical methods that preserve these conservation laws.
Existing Casimir preserving methods, however, rely on computationally expensive mappings between Lie algebras and Lie groups, such as the exponential map, which render them impractical for large-scale applications.
In this work, based on recent advances in geometric integration \cite{Modin2020,Modin2020_sphere}, we show how to devise an efficient Casimir-preserving method, exploiting optimised linear algebra algorithms in conjunction with scalable parallel computing techniques, that allows simulation of high-resolution flows. 

The Hamiltonian description of two-dimensional ideal hydrodynamics carry the structure of an infinite-dimensional \emph{Lie-Poisson system} (cf.~\cite{Marsden_book}).
The first step is to find a finite-dimensional Lie-Poisson system corresponding to a spatial discretization of the original PDE.
This was achieved by Zeitlin \cite{Zeitlin1991,Zeitlin2004}, who provided a self-consistent truncation of the Euler equations based on the theory of geometric quantisation, as devised by Hoppe \cite{Hoppe1989} and later analysed by Bordemann et.\ al.~\cite{Bordemann1991,Bordemann1994}.
Quantisation, here, refers to the process of constructing a sequence of matrix Lie algebras for which the Poisson bracket is approximated by the matrix commutator.
Once such a construction is accomplished, a Lie-Poisson time-step integrator should be employed to preserve the underlying geometric structure of the system.

The finite-mode approximation of hydrodynamics proposed by Zeitlin has been investigated through numerical simulation. Indeed, McLachlan \cite{McLachlan1993} developed a Lie-Poisson preserving integrator for simulation of Euler's equations on the flat torus. 
This method exploits a number of symmetries of the quantised Euler equations on the periodic square to devise a clever splitting of the flow map. 
Each component of the flow map is, in fact, a Poisson map that is solved exactly by means of the fast Fourier transform. 
The complexity of McLachlan's algorithm is $\mathcal{O}(N^3 \ln N)$ for $N^2$ spatial degrees of freedom. 
In comparison, the direct approach requires $\mathcal{O}(N^4)$ computations only to evaluate the quantised Poisson bracket. 
Nevertheless, even though the single map components can be solved efficiently, $\mathcal{O}(N)$ such maps have to be computed sequentially which render the overall method unfeasible for $N \gtrsim 10^2$. 

Since the work of McLachlan, little progress was made on algorithms based on Zeitlin's model until the work of Modin and Viviani \cite{Modin2020,Modin2020_sphere}, which provides a new class of Lie-Poisson integrators for isospectral flows.
The governing equations are cast into an evolution equation for the vorticity matrix by making use of the quantised basis derived by Hoppe~\cite{Hoppe1989}.
The resulting discrete system is isospectral, i.e., the eigenvalues of the vorticity matrix are constants of motion.
In turn, conservation of the spectrum is analogous to conservation of the Casimir functions.
Specifically, the methods in \cite{Modin2020} are based on a discrete Lie-Poisson reduction of symplectic Runge-Kutta schemes, where time-stepping is performed directly at the level of the matrix Lie algebra to avoid computationally unfeasible Lie algebra to Lie group maps.
The new isospectral methods thereby presents a potential for the development of efficient algorithms for geometric integration of systems characterised by a large number of degrees of freedom such as fluid systems: the subject of our work here.

The complexity of the isospectral methods by Modin and Viviani~\cite{Modin2020} is $\mathcal{O}(N^3)$ per time-step due to matrix multiplication in the commutator. An optimal method would have complexity $\mathcal{O}(N^2)$, thus scaling linearly with the spatial degrees of freedom. 
To construct a Lie-Poisson integrator with complexity lower than $\mathcal{O}(N^3)$ is still an open problem, so an algorithm with optimal complexity is currently out of reach.
Nevertheless, the isospectral integrator developed here is characterised by a low count of matrix-matrix multiplications per time-step and for those multiplications we use highly optimised and scalable linear-algebra packages. Furthermore, the block-tridiagonal form of the discrete Laplacian can be efficiently exploited to compute the quantised basis and the stream function. 
This technique allows for simulations of matrix size $N\gtrsim 10^3$ allowing for the resolutions required for the wide spectrum of scales of motion typical of fluid systems, such as turbulent flows.

Although the geometric integrator we develop is applicable to different fluid domains, we focus here on flows on the sphere. In contrast to the flat torus---the domain most often simulated in numerical studies---the sphere is more appealing from a geophysical viewpoint. Moreover, no artificial boundary conditions have to be imposed making the sphere an ideal numerical test-ground \cite{Lindborg2022}.

The paper is structured as follows. 
In Sec.~\ref{sec:num} the developed Lie-Poisson integrator is presented, via the quantised matrix formulation of Euler's equations as a spatial discretisation scheme.
In Sec.~\ref{sec:par} the details of the numerical algorithms and their parallelisation are discussed.
The capability of the method developed is illustrated in Sec.~\ref{sec:appl}, where a long-time simulation of the Euler equations on the unit sphere is presented. 
Flow visualisations and the kinetic energy spectrum at different times are also given. 
Conclusions and main results are given in Sec.~\ref{sec:con}. 

\section{Problem definition and numerical method} \label{sec:num}
In this section we begin by summarising the main steps that lead to the construction of the Lie-Poisson integrator developed in \cite{Modin2020_sphere}. 
To this end, consider the incompressible Euler equations on the unit sphere $\mathbb{S}^2\subset \mathbb{R}^3$ embedded in Euclidean 3-space. 
Expressed in the vorticity $\omega\colon \mathbb{S}^2\to \mathbb{R}$, they are
\begin{equation}
\begin{cases}
\dot{\omega} = \{ \psi,\omega \}, \\
\Delta \psi = \omega,
\end{cases}
\label{eq:Euler_vort}
\end{equation} 
where $\psi$ is the stream-function, related to vorticity $\omega$ via the Laplace-Beltrami operator $\Delta$, and $\{\cdot,\cdot \}$ is the Poisson bracket defined by
\[
\{ \psi,\omega \}(x)=x\cdot(\nabla\psi\times\nabla\omega), \qquad x\in\mathbb{S}^2\subset\mathbb{R}^3.
\]
% for any $x\in\mathbb{S}^2$, $\psi$ the stream function and $\Delta$ the Laplace-Beltrami operator. 
The Euler equations~\eqref{eq:Euler_vort} constitute a Lie--Poisson system on $C^{\infty} (\mathbb{S}^2)$, for the Lie--Poisson bracket given by
\[
\lbrace \mathcal{F}, \mathcal{G}\rbrace_{LP}(\omega)=\int_{\mathbb{S}^2}\omega \left\{ \frac{\delta\mathcal{F}}{\delta\omega}(\omega),\frac{\delta\mathcal{G}}{\delta\omega}(\omega) \right\},\]
for smooth functionals $\mathcal{F,G} \colon C^{\infty} (\mathbb{S}^2)\to \mathbb{R}$.
With respect to this Lie--Poisson structure, the Hamiltonian that yields the Euler equations~\eqref{eq:Euler_vort} is
\[
\mathcal{H}(\omega)=-\dfrac{1}{2}\int \omega\psi.
\]
In other words, equation \eqref{eq:Euler_vort} can be written
\begin{equation}\label{eq:Euler_vort_LP}
\dfrac{d\mathcal{F}}{dt}(\omega)=\lbrace \mathcal{F}, \mathcal{H}\rbrace_{LP}(\omega),\qquad \forall \mathcal{F} : C^{\infty} (\mathbb{S}^2)\rightarrow \mathbb{R}.
\end{equation}
% $$.
For this system there exists infinitely many Casimir functions, i.e., functionals $\mathcal{C}: C^{\infty} (\mathbb{S}^2)\rightarrow \mathbb{R}$ such that
\begin{equation}
    \begin{aligned}    
    \{ \mathcal{F},\mathcal{C} \}_{LP} \equiv 0 && \forall \mathcal{F} : C^{\infty} (\mathbb{S}^2)\rightarrow \mathbb{R}.
    \label{eq:Casimir_bracket}
    \end{aligned}
\end{equation}
In particular, the integrated powers of vorticity
\begin{equation}
\begin{aligned}
\mathcal{C}_k(\omega) = \int \omega^k, && k=1,2,\ldots
\end{aligned}
\label{eq:Casimirs}
\end{equation}
are invariants of motion independently of the choice of Hamiltonian $\mathcal {H}$.

In order to derive a discretisation that captures the conservation laws (\ref{eq:Casimir_bracket}) in a discrete sense, it is essential to look at Euler's equations from the geometric viewpoint and embed the underlying differential structure into the discrete system. This process of consistent discretisation is called \emph{quantisation} (cf.~\cite{Hoppe1989,Bordemann1991,Bordemann1994}). 
For completeness, we briefly summarise its meaning in the following.

As a first step, the construction of a numerical scheme that embeds a Lie-Poisson structure requires a finite truncation of the Poisson bracket. 
In \cite{Zeitlin1991,Zeitlin2004}, Zeitlin gave an approach based on the theory of geometric quantisation. 
This consists of a sequence of matrix Lie algebras of real dimension $N^2$, which approximates the infinite-dimensional Poisson algebra of functions as $N$ goes to infinity. 
Unrelated to the work of Zeitlin, but motivated by natural questions of classical limits in theoretical physics, Bordemann et.~al.~\cite{Bordemann1994} showed that there exists a basis of $\mathfrak{u}(N)$ (the Lie algebra of skew-Hermitian matrices) for which the structure constants converge to those of the spherical harmonics basis of the Poisson algebra $\mathcal{C}^{\infty}(\mathbb{S}^2)$. 
A projection $\Pi_N \colon \mathcal{C}^{\infty}(\mathbb{S}^2) \to \mathfrak{u}(N)$ can be constructed, via the spherical harmonics basis $Y_{lm}$ for $C^\infty(\mathbb{S}^2)$ and a corresponding basis $T^N_{lm}$ for $\mathfrak{u}(N)$, such that for $N \to \infty$
\begin{equation}
\begin{aligned}
& \Pi_N f - \Pi_Ng \to 0 \ \text{implies} \ f=g, \\
&\Pi_N \{ f,g \} = \frac{N^{3/2}}{\sqrt{16\pi}}[\Pi_Nf, \Pi_Ng] + \mathcal{O}(1/N^2).
\end{aligned}
\label{eq:cond_commutator}
\end{equation}
In other words, we have a way of approximating smooth functions on $\mathbb{S}^2$ by finite-dimensional matrices in $\mathfrak{u}(N)$: this is the essence of Zeitlin's idea.
The setting naturally restricts to the subalgebra $\mathfrak{su}(N)\subset \mathfrak{u}(N)$ of traceless matrices, which correspond to zero mean vorticity and stream-function. 
Using the discrete basis $T^N_{lm}$ one then obtains  a spatial discretization of (\ref{eq:Euler_vort}) in the matrix form \cite{Zeitlin2004,Modin2020}:
\begin{equation}
\begin{cases}
\dot{W} = [P,W]\\
\Delta_N P = W,
\end{cases}
\label{eq:Euler_vort_mat}
\end{equation}
where $W  \in \mathfrak{su}(N)$ is the \emph{vorticity matrix}, $P  \in \mathfrak{su}(N)$ is the \emph{stream matrix} and $\Delta_N$ is the quantized Laplacian given by \cite{Hoppe1998}. 
Notice that traces of powers of the vorticity matrix~$W$,
\begin{equation}
\begin{aligned}
C_k(W)=\text{Tr}(W^k) && \text{for } k=1,\ldots,N,
\end{aligned}
\label{eq:W_pow}
\end{equation}
are conserved by the system (\ref{eq:Euler_vort_mat}). 
This is the discrete analogue of conservation of integrated powers of vorticity in the continuum. 

Once the spatial discretization~\eqref{eq:Euler_vort_mat} is obtained, the second step is to derive a time-stepping integrator such that the Casimir functions (\ref{eq:W_pow}) are conserved. 
The key observation is that the matrix flow~\eqref{eq:Euler_vort_mat} is isospectral (it preserves the eigenvalues of $W$), which in turn implies conservation of Casimirs.
(More specifically, isospectrality is equivalent to preservation of \emph{co-adjoint orbits}, cf.~\cite{Marsden_book}.)
% when the {\color{red}time derivative of a skew-symmetric matrix, $W$, is equal to the commutator of $W$ with another skew-symmetric matrix, then the flow is isospectral, i.e. the eigenvalues of $W$ are constant over time \cite{Hairer_book}}. In turn, conservation of the spectrum of $W$ is analogous to conservation of the Casimirs. 
Now, the methods by Modin and Viviani~\cite{Modin2020} (see also Viviani~\cite{Viviani2020}) are precisely applicable for Lie-Poisson systems realised as isospectral flows. 
The approach uses symplectic Runge-Kutta methods in conjunction with discrete Lie--Poisson reduction to derive numerical schemes for which the time-stepping is performed directly at the level of the matrix Lie algebra. 
The approach avoids usage of the exponential map.
% , which would otherwise make geometric integration unfeasible in terms of computational cost for large $N$. 
A particularly simple integrator of this class, and the one we adopt here, arises from the symplectic midpoint method and is given as follows \cite{Viviani2020}:
\begin{equation}
\begin{cases}
W_n = \left( I-\frac{h}{2} \Delta_N^{-1} \widetilde{W} \right) \widetilde{W} \left( I+\frac{h}{2} \Delta_N^{-1} \widetilde{W} \right) \\
W_{n+1} = \left( I+\frac{h}{2} \Delta_N^{-1} \widetilde{W} \right)\widetilde{W} \left( I-\frac{h}{2} \Delta_N^{-1} \widetilde{W} \right),
\end{cases}
\label{eq:isomp}
\end{equation}
with $n$ the time level, $h$ the time-step and $\widetilde{W}$ an implicitly defined matrix. The computation of the latter requires an iterative method. 
We found the following simple fixed point iteration to be efficient \cite{Benzi2021}:
\begin{equation}
\widetilde{W}^{k+1} = W_n + \frac{h}{2} \left( \Delta_N^{-1} \widetilde{W}^k \widetilde{W}^k  - \widetilde{W}^k  \Delta_N^{-1} \widetilde{W}^k \right) + \frac{h^2}{4} \Delta_N^{-1} \widetilde{W}^k \widetilde{W}^k  \Delta_N^{-1}  \widetilde{W}^k,
\label{eq:isomp_iter}
\end{equation}
where $k$ indicates the iteration step. 
Integration of (\ref{eq:Euler_vort_mat}) by means of (\ref{eq:isomp}) preserves the spectrum of $W$, and thus the discrete Casimir functions, up to the tolerance set for the convergence of (\ref{eq:isomp_iter}). 
Numerical tests have shown that a number of $2$ to $3$ iterations are sufficient to reach a tolerance of $10^{-12}$ in the infinity norm of the matrix $(\widetilde W^{k+1}-\widetilde W^k)$. This is due to the fact that the initial vorticity is always normalized (with respect to the spectral norm) and that the scaling factor $N^{3/2}$ of the commutator in (\ref{eq:cond_commutator}) is absorbed in the time-step (see \cite{Benzi2021} for a more extensive discussion on the relation between the number of iterations with respect to $N$). 

In the next section we present a detailed account of the numerical algorithms employed for the solution of (\ref{eq:isomp}) and their parallelisation. 

\section{Main algorithms and parallelisation} \label{sec:par}
The numerical scheme (\ref{eq:isomp}) can be broken down into three main components:
\begin{itemize}
    \item the computation of matrix multiplications stemming from the commutator (Sec. \ref{sec:mat_comm}), 
    \item the construction the quantised basis (Sec. \ref{sec:q_basis}), and 
    \item the computation of the inverse Laplacian (Sec. \ref{sec:inv_lap}).
\end{itemize}
Linear algebra algorithms are taken from the well-established and optimised library LAPACK \cite{lapack} and its parallel extension ScaLAPACK \cite{scalapack}. 
The parallelisation will be carried out by means of MPI \cite{Gropp1996} combined with openMP multithreading \cite{chandra2001parallel}. 
The overall complexity of the developed Lie-Poisson integrator is reduced to that of matrix multiplication resulting from the commutator, the latter being $\mathcal{O}(N^3)$. 
Therefore, we select a distribution memory layout that allows for optimal computation of matrix-matrix multiplications referred to as \textit{block-cyclic decomposition} \cite{scalapack}. 
In essence, the latter assigns matrix blocks to MPI processes in a cyclic manner in order to optimise load-balance and communication across processors for dense matrix operations. 
We refer to the ScaLAPACK User' Guide for a complete description. 

\subsection{Computation of the matrix commutator} \label{sec:mat_comm}
The commutator in the quantised system (\ref{eq:Euler_vort_mat}) gives rise to matrix multiplications in the isospectral integrator (\ref{eq:isomp}). 
Dense matrix multiplications are carried out by the routine \texttt{pzgemm}, which optimises data communication across MPI processes through a block-cyclic decomposition of the computational domain \cite{scalapack}. Inspection of (\ref{eq:isomp_iter}) shows that three matrix multiplications are needed for the computation of each iteration of $\widetilde{W}^k$. However, given the skew-symmetry of both the stream matrix and the vorticity matrix, the third term on the right-hand-side of (\ref{eq:isomp_iter}) can be computed as the conjugate transpose of the second term. Clearly, the same simplification applies also to the second equation in (\ref{eq:isomp}). Thus, only two dense matrix multiplications, of complexity $\mathcal{O}(N^3)$, are needed for each update of $\widetilde{W}^k$ and $W_{n+1}$.  

\subsection{Computation of the quantised basis} \label{sec:q_basis}

Given an initial condition in terms of spherical harmonic coefficients, $\omega_{lm}$, the vorticity matrix is assembled by the projection onto the quantised basis
\begin{equation}
W_0 = \sum_{l=0}^{N-1} \sum_{m=-l}^l \mathrm{i}\omega_{lm} \widehat{T}^N_{lm},
\label{eq:W0}
\end{equation}
where 
\begin{align*}
    T^N_{lm}&=\mathrm{i}((\widehat{T}^N_{lm} + (-1)^m(\widehat{T}^N_{l-m})/\sqrt{2},\mbox{ for }m>0 \\
    T^N_{lm}&= (\widehat{T}^N_{lm} - (-1)^m\widehat{T}^N_{l-m})/\sqrt{2},\mbox{ for }m<0 \\
    T^N_{l0} &= \mathrm{i}\widehat{T}^N_{l0},\mbox{ for }m=0,
\end{align*}
for the basis elements $T^N_{lm} \in \mathfrak{su}(N)$ introduced in Sec. \ref{sec:num}. In \cite{Hoppe1989} an explicit expression for the computation of $\widehat{T}^N_{lm}$ is given:
\begin{equation}
\left(\widehat{T}^N_{lm}\right)_{ij} = (-1)^{ [ (N-1)/2 ]-i } \sqrt{2l+1} 
\begin{pmatrix}
\frac{N-1}{2} & l & \frac{N-1}{2}\\
-i & m & j
\end{pmatrix},
\label{eq:Tlm}
\end{equation}
where the bracket denotes the Wigner 3j-symbols. While the analytical expression (\ref{eq:Tlm}) is theoretically appealing, it is in practice unfeasible to compute as $N$ becomes large. 
This is due to the fact that known analytic formula for the 3j-symbols contain large factorials that are computationally expensive and difficult to compute on a computer. Here, we suggest a more efficient alternative, essential for numerical simulation of fluid systems. 
Analogous to the continuum case, the discrete spherical Laplacian satisfies \cite{Hoppe1998}
\begin{equation}
\begin{aligned}
\Delta_N \widehat{T}^N_{lm} = -l(l+1) \widehat{T}^N_{lm}, && \forall \ l=1,...,N, m=-l,...,l.
\end{aligned}
\label{eq:lap_prop}
\end{equation}
Thus, matrices $\widehat{T}^{N}_{lm}$ can be found as solutions to the eigenvalue problem (\ref{eq:lap_prop}). The discrete Laplacian is a forth-order tensor, hence dense linear algebra would require $\mathcal{O}(N^4)$ operations to apply the quantised Laplacian. 
However, $\Delta_N$ admits splitting into $(2N-1)$ blocks $\Delta^m$ of size $N-|m|$ corresponding to the $m$th diagonal of the matrix $W$. 
In particular, for $m \geq 0$ one has
\begin{equation}
\begin{aligned}
\Delta^m_{ij} = & 2\delta_{i}^j \left( s(2i+1+m) - i(i+m) \right) \\
& - \delta_{i+1}^j \sqrt{(i+m+1)(N-1-i-m)}\sqrt{(i+1)(N-1-i)} \\
& - \delta_{i-1}^j \sqrt{(i+m)(N-i-m)}\sqrt{i(N-i)},
\end{aligned}
\label{eq:delta_tria}
\end{equation}
where $s=(N-1)/2$ and $i,j=0,...,N-m-1$. 
Thus, not only does the Laplacian split into blocks corresponding to the diagonals, but also on each such block the operator $\Delta^m$ is sparse corresponding to a tribanded matrix.
The quantised basis can hence be computed by solving $N$ tridiagonal eigenvalue problems at most of size $N$. 
The computation of the eigenvectors is carried out by the parallel LAPACK routine \texttt{pdstedc} \cite{scalapack}, a highly optimised library for symmetric tridiagonal matrices which enables data distribution through MPI \cite{Gropp1996}. The remaining components of (\ref{eq:W0}) for $m<0$ can be found by skew-symmetry. Formulated as an eigenvalue problem, the overall complexity of the computation of the quantised basis is then $\mathcal{O}(N^3)$. 
We remark that the elements of the basis are only needed to generate the initial conditions and at a few selected times when the spherical harmonic coefficients $\omega_{lm}$ are extracted from $W$ for post-processing.

\subsection{Computation of the inverse Laplacian} \label{sec:inv_lap}
Scheme (\ref{eq:isomp}) involves the computation of the stream matrix $P=\Delta^{-1}_N W$. The discrete Laplacian $\Delta_N$ is a fourth-order tensor, which can be expressed as a matrix of size $N^2\times N^2$. 
As such, a naive computation of its inverse would be unfeasible. 
However, $\Delta_N$ splits into the blocks (\ref{eq:delta_tria}) and on each such block we have a tridiagonal system. 
Thus, the problem of finding the stream matrix can be formulated in terms of a set tridiagonal systems, one on each block. 
To this aim we introduce the following notation. 
Given an $N\times N$ matrix $A$, we identify $A_m$ with its $m$-th diagonal defined as the $m$-th sub-diagonal for $1\leq m \leq N-1$ and the main diagonal for $m=0$. 
As shown in \cite{Hoppe1998}, the tridiagonal Laplacian (\ref{eq:delta_tria}) acts on stream-matrix diagonal elements $P_m$ and produces vorticity-matrix diagonal elements $W_m$, i.e.,
\begin{equation}
\begin{aligned}
{ \Delta^m P_m = W_m}, && {m=0,..,N-1.}
%P[i=m+1,...,N;j=i-m] = \left( \Delta^m \right)^{-1} W[i=m+1,...,N;j=i-m], && m=0,..,N-1,
\end{aligned}
\label{eq:sub_tria_W}
\end{equation}
The solution of systems (\ref{eq:sub_tria_W}) yields the lower triangular part of $P$. Moreover, since $P \in \mathfrak{su}(N)$, its upper triangular part is computed by skew-symmetry.
In summary, the stream matrix can be computed by solving $N$ tridiagonal systems, the largest (for $m=0$) having size $N$. 
Thus, the overall complexity to solve for the stream matrix is reduced to $\mathcal{O}(N^2)$. 

\begin{figure}[hbt!]
\centering
\includegraphics[width=0.5\textwidth]{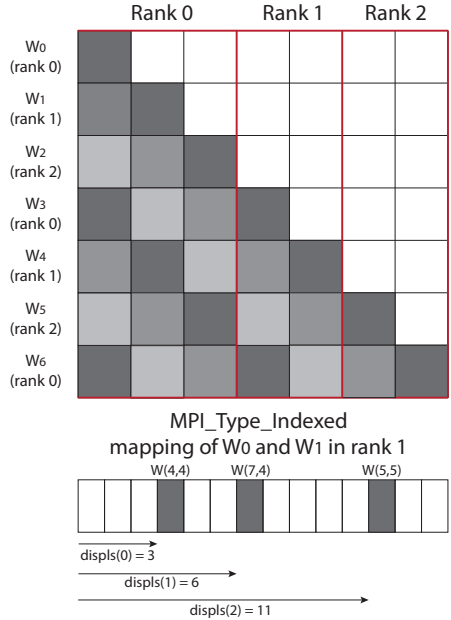}
\caption{Sketch of a matrix of size $N=7$ distributed over three MPI processes across columns. The red boxes indicate the data local to each MPI rank. The diagonals $W_m$ of the vorticity matrix, corresponding to the tridiagonal systems (\ref{eq:sub_tria_W}), are represented by filled blocks. $W_m$ having the same color are assigned to the same MPI rank, as indicated in the left of the figure. The construction of the \texttt{MPI\_Type\_Indexed} related to the mapping of $W_0$ and $W_3$ stored in the memory of rank $1$ is shown at the bottom of the figure. The diagonal values are identified by specifying their displacement in memory as indicated by arrows at the bottom of the figure.}
\label{fig:mpi_type}
\end{figure}

Particular care has to be taken when implementing (\ref{eq:sub_tria_W}) on a distributed-memory system. As pointed out in Sec. \ref{sec:q_basis}, the matrix data layout is based on a block-cycling distribution, which is optimal for matrix multiplication. Therefore, the $N$ diagonals of $W$ are scattered among processors and mapped into memory in a non-trivial way. In order to extract the diagonals in a simple and efficient way, we first redistribute $W$ into a block-column memory layout (see Fig. \ref{fig:mpi_type}) via \texttt{pzgemr2d}, a routine that provides a copy among different matrix memory layouts within the ScaLAPACK standard. Subsequently, an approximately equal number of diagonals of $W$ is assigned to the available MPI ranks. As it is clear from Fig. \ref{fig:mpi_type}, some values of the diagonals owned by a given MPI rank are stored in the local memory of a different MPI rank. This array of values is communicated among processors using derived types \texttt{MPI\_Type\_Indexed}. The latter are a particularly efficient means of parallel communication when the data structure of the algorithm remains constant throughout its execution, as it is the case here. One can, in fact, encode the memory layout of the data to be transferred into the derived data type and send them across processors in a single MPI instruction, thus reducing communication to a minimum and avoiding buffering of data altogether. Furthermore, data are transferred in a non-blocking manner, by the pair \texttt{MPI\_Issend}, \texttt{MPI\_Irecv}, reducing waiting times.

To illustrate the parallel strategy for the computation of the stream matrix and the construction of derived types \texttt{MPI\_Type\_Indexed}, we sketch the data transfer sent by rank $1$ to rank $0$, for a $7 \times 7$ matrix distributed over three MPI processes (Fig. \ref{fig:mpi_type}). 
Rank $0$ handles the tridiagonal systems (\ref{eq:sub_tria_W}) for $m=0,3,6$, whose right-hand sides $W_m$ are indicated by the dark grey blocks. 
Clearly, part of $W_0$ and $W_3$ is stored in the local memory of rank $1$ and needs to be communicated to rank $0$ for the systems to be solved. 
These matrix elements, $W(4,4)$, $W(7,4)$ and $W(5,5)$, are first mapped into an indexed data type by specifying their memory displacement from the first element held in memory by rank $1$, i.e., from the element $W(1,4)$ following a column-major order (see sketch at the bottom of Fig.~\ref{fig:mpi_type}).
Subsequently, the mapped values are sent with a single MPI instruction to rank $0$ where they are reconstructed to form contiguous buffers containing the full right-hand sides $W_m$. 
The corresponding set of tridiagonal systems is then solved, by a pool of threads, employing the LAPACK routine \texttt{zpttrs}, a dedicated linear solver for Hermitian positive definite tridiagonal matrices. 
Once the solution vectors are computed, they are sent back to their corresponding diagonals $P^m$ and, finally, the matrix $W$ is redistributed into the block-cycling memory layout. 
This procedure is applied by all MPI ranks.

Scaling tests indicate that the developed parallel algorithm is efficient and scalable, as we illustrate in the next section.

\subsection{Parallel performance} \label{sec:par_perf}
The parallel performance of the algorithm detailed in Sec.~\ref{sec:par} is analysed on the supercomputer Galileo100~\cite{Galileo100}, which mounts Intel CPU CascadeLake 8260 equipped with $24$ cores each. We carry out a scaling test for matrix size $N=2048$ and $N=4096$. 
These resolutions allow for the study of complex flows that span a wide range of scales of motion, as demonstrated in \cite{Cifani2022} for the case of two-dimensional turbulence. 
The number of iterations in (\ref{eq:isomp_iter}) is set to~$3$. As pointed out in Sec.~\ref{sec:num}, this was found to be sufficient to reach a tolerance of $10^{-12}$, independently of $N$, for a time-step that resolve the flow time-scale.

\begin{figure}[hbt!]
\centering
\begin{subfigure}[b]{0.45\textwidth}
\includegraphics[width=1\textwidth]{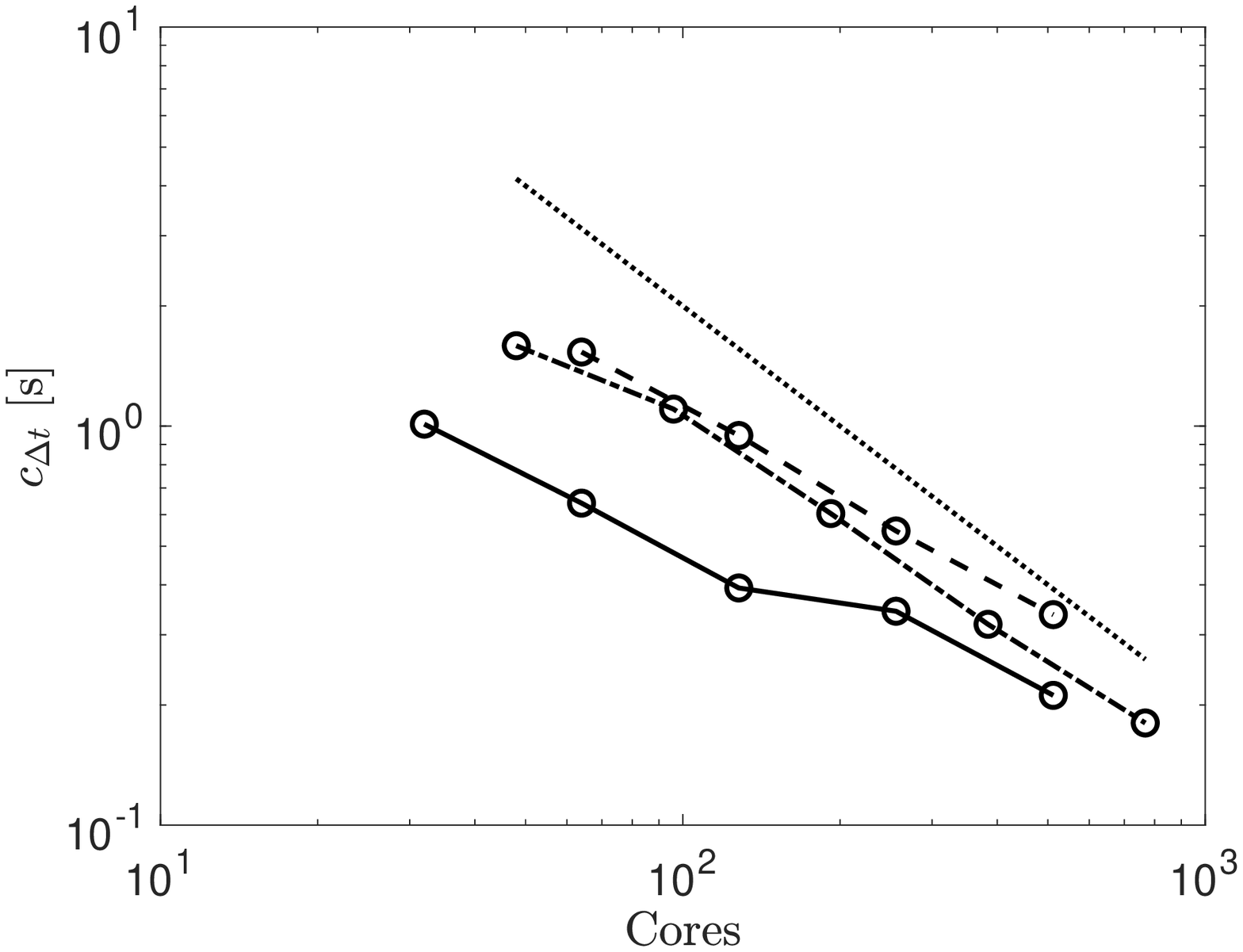}
\end{subfigure}
\hfill
\begin{subfigure}[b]{0.45\textwidth}
\includegraphics[width=\textwidth]{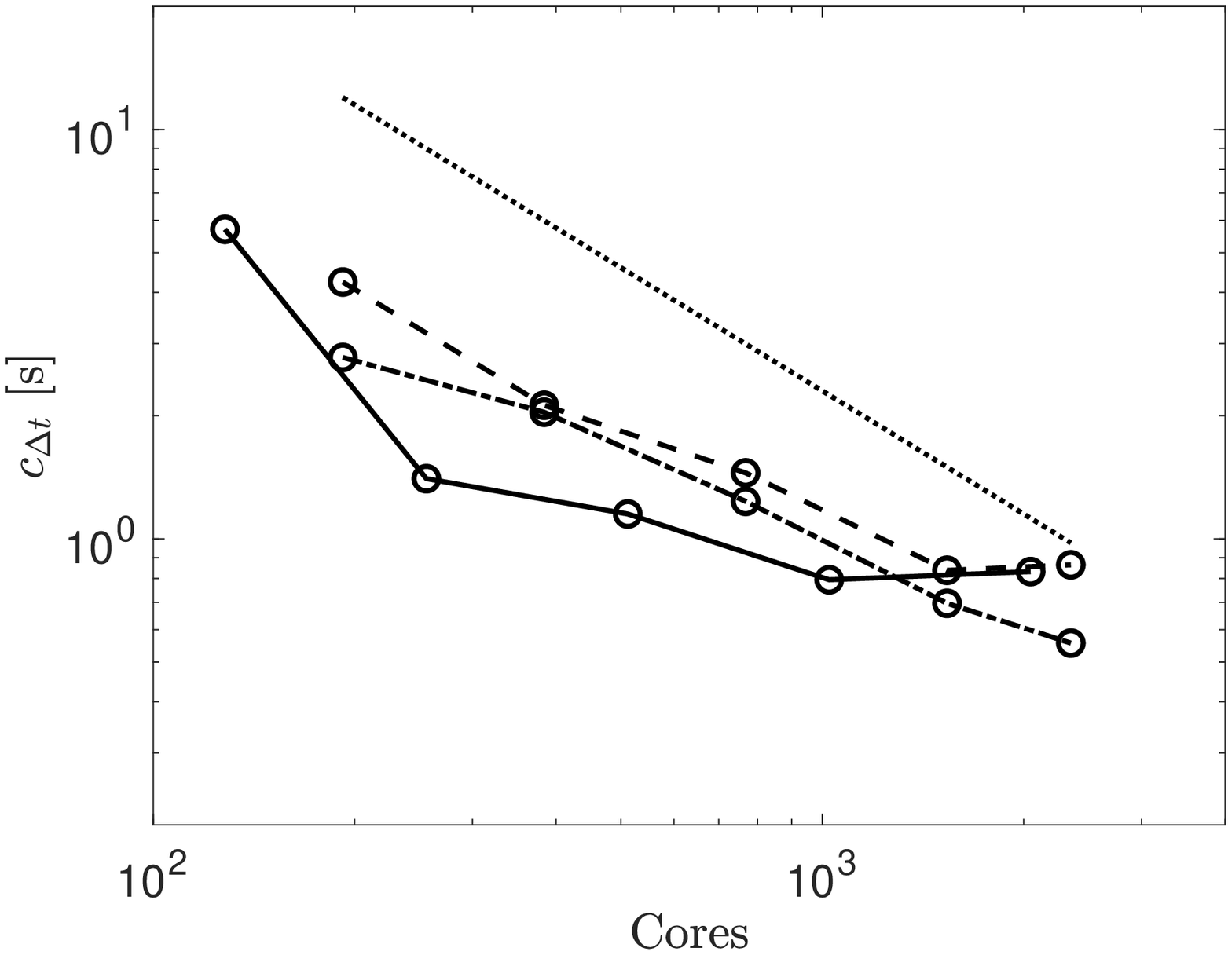}
\end{subfigure}
\caption{Computational time per time-step as a function of the number of cores, for $N=2048$ (left figure) and for $N=4096$ (right figure), using fully MPI parallelisation (solid line), hybrid MPI parallelisation with $12$ threads (dash-dotted line) and hybrid MPI parallelisation with $24$ threads (dashed line). Linear scaling is shown as a reference by the dotted line.}
\label{fig:scaling}
\end{figure}

Fig.~\ref{fig:scaling} shows the computational time per time-step as a function of the number of cores. 
A full MPI parallelisation (solid line) is compared with an hybrid parallelisation for two different number of threads per MPI process equal to $12$ (dash-dotted line) and $24$ (dashed line). 
The best performance is found when employing $12$ threads: approximately linear scaling is observed with the lowest computational time at the largest number of cores simulated. 
The full MPI parallelisation appears to be more efficient for smaller matrix sizes, while it deviates from linear scaling for large $N$. 
Doubling the number of threads to $24$ does not improve the computational time and results in an earlier departure from linear scaling compared to the case where $12$ threads are employed. 
Our findings thus indicate that a certain degree of multithreading leads to an overall better parallel performance for large $N$. 
However, we remark that the degree at which hybrid parallelisation is favourable is system-dependent. 
What ultimately matters is the computational time one can achieve by means of parallel computing. 
Here we show that, for $N=4096$, a time-step is completed in around $0.55$ seconds, which in turn allows for long-time simulations and gathering of statistical quantities of the flow, as for example demonstrated in \cite{Cifani2018} and in the following section.

\section{Simulation of Euler's equations} \label{sec:appl}
\begin{figure}[hbt!]
\centering
\begin{subfigure}[b]{0.45\textwidth}
\includegraphics[width=1\textwidth]{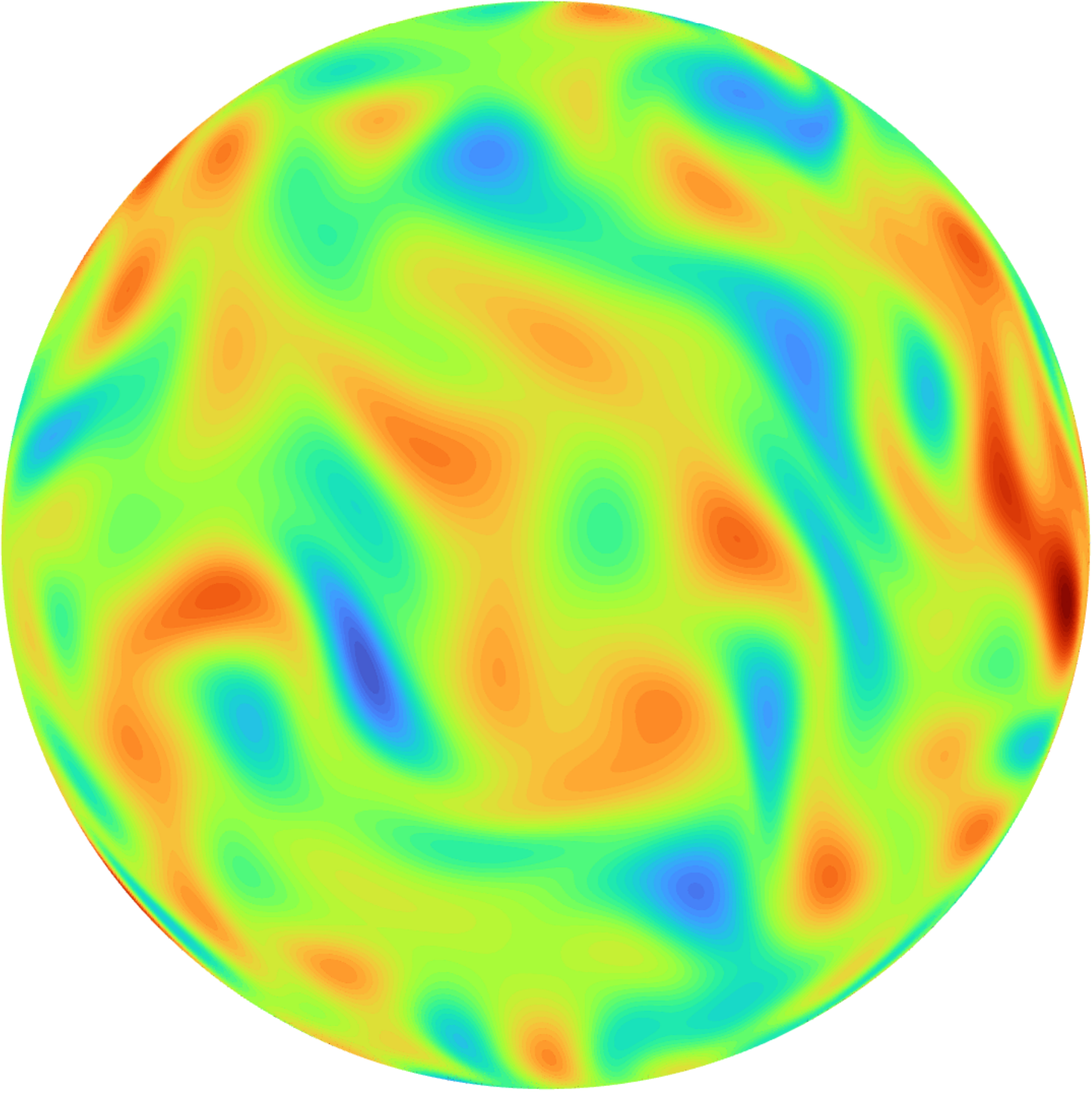}\\
\end{subfigure}
\hfill
\begin{subfigure}[b]{0.45\textwidth}
\includegraphics[width=\textwidth]{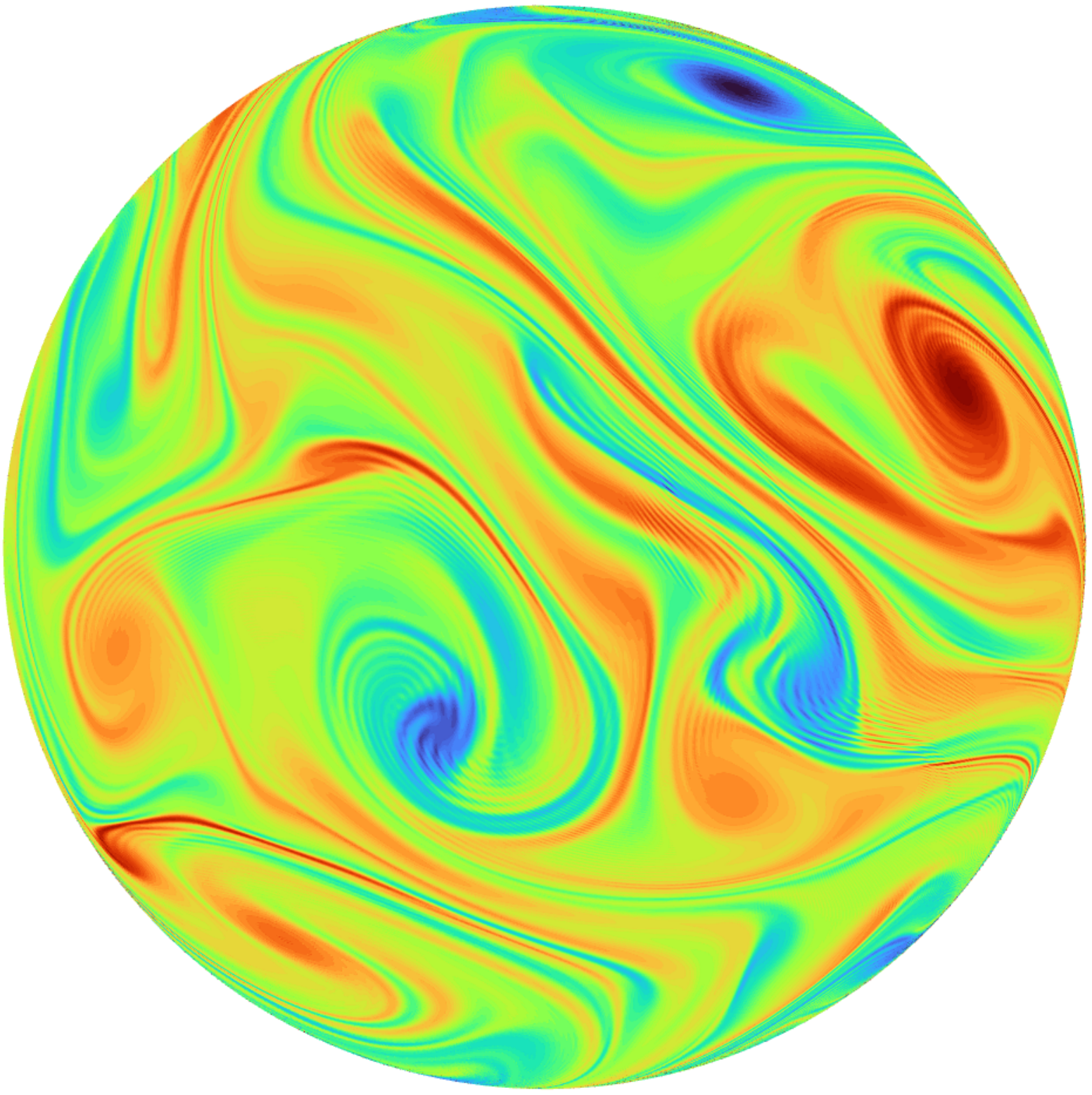}\\
\end{subfigure}
\begin{subfigure}[b]{0.45\textwidth}
\includegraphics[width=\textwidth]{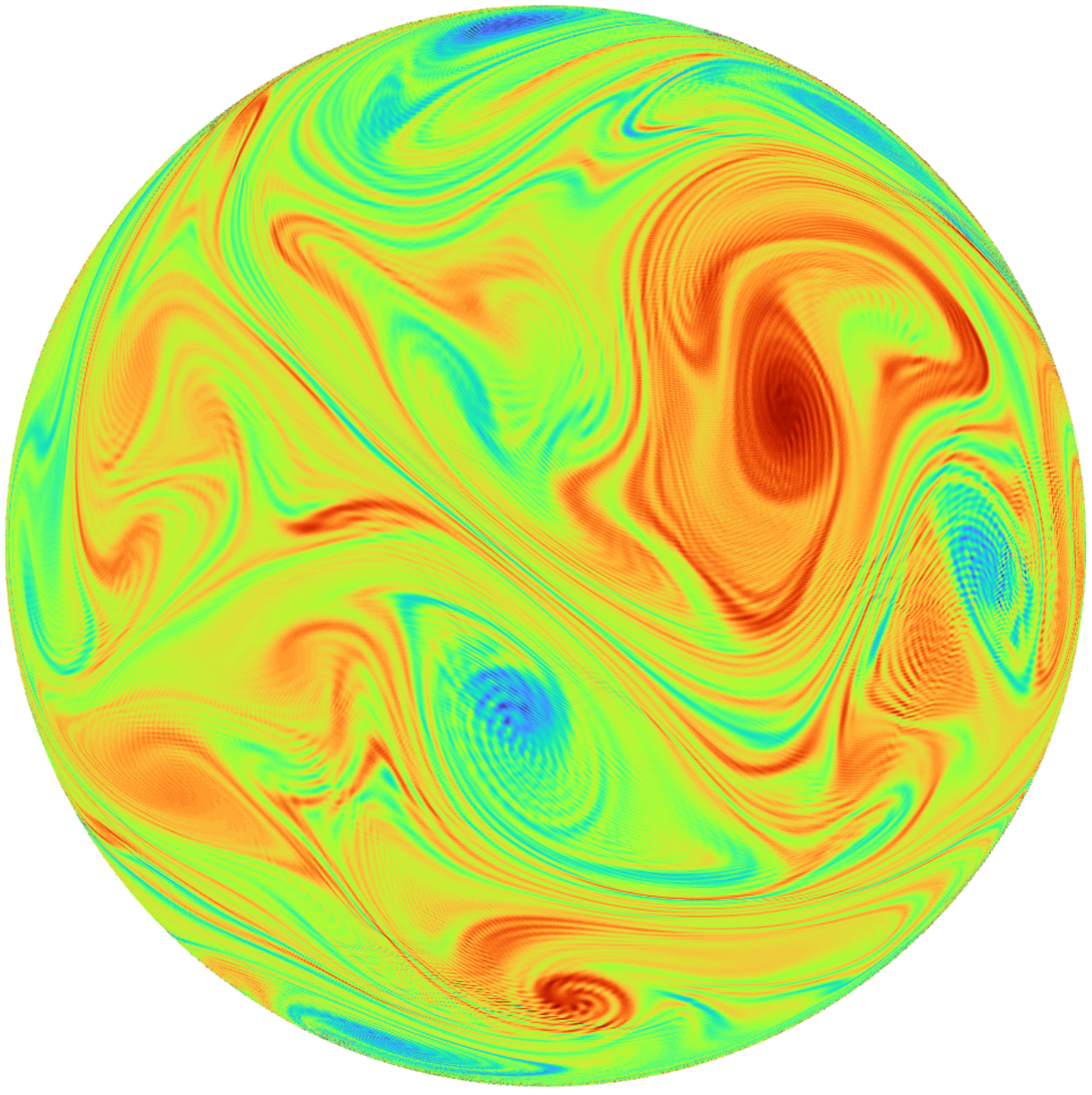}\\
\end{subfigure}
\hfill
\begin{subfigure}[b]{0.45\textwidth}
\includegraphics[width=\textwidth]{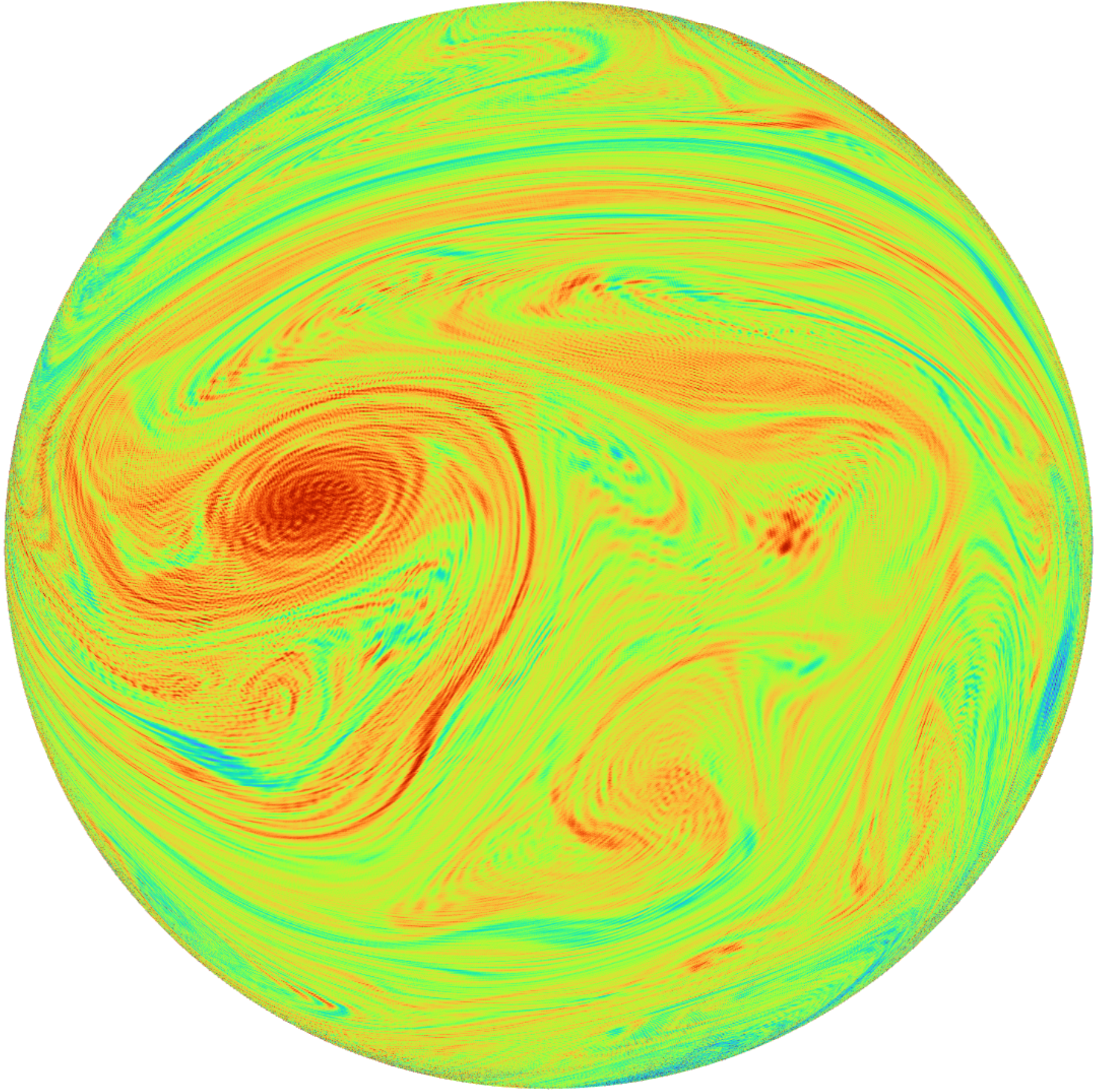}\\
\end{subfigure}
\caption{Vorticity field at $t=10$ (top-left figure), $t=60$ (top-right figure), $t=100$ (bottom-left figure) and $t=200$ (bottom-right figure). Colours range from blue corresponding to $\omega=-0.8$ to red corresponding to $\omega=0.8$.}
\label{fig:vort_snaps}
\end{figure}

In this section we illustrate the capability of the developed numerical method by simulating the discrete Euler equations (\ref{eq:Euler_vort_mat}) for $N=2048$. At the initial time, modes of spherical harmonics of degree $2 \leq l \leq 20$ are given a random value of approximately the same magnitude. The system then evolves for a total of $10^6$ time-steps corresponding to $200$ time units. For this simulation $512$ cores were employed. A qualitative illustration of the evolution of the vorticity field is shown in Fig. \ref{fig:vort_snaps}. 
From an initial random distribution of vorticity in spherical harmonic space, the flow evolves into well recognisable large-scale vortical structures. At the same time, owing to non-linear interactions, part of the energy injected at larger scales flows toward the high-wave-number end of the spectrum. Since the simulated flow is inviscid and the geometric integrator preserves the integrated powers of vorticity, neither physical dissipation nor artificial numerical dissipation play a role in the flow dynamics. As a result, after a sufficiently long time all simulated scales will acquire a certain amount of energy, as it becomes visible by the roughness of the vorticity field at $t=200$. 

\begin{figure}[hbt!]
\centering
\includegraphics[width=0.6\textwidth]{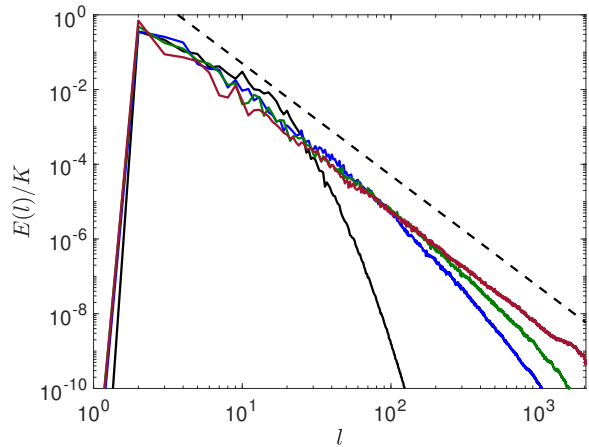}
\caption{Kinetic energy spectrum scaled by the mean kinetic energy $K$ as a function of the spherical harmonic index $l$ at different times: $t=10$ (black line), $t=60$ (blue line), $t=100$ (green line) and $t=200$ (red line). The $l^{-3}$ scaling is represented by the dashed black line.}
\label{fig:spectra_Euler}
\end{figure}

In Fig. \ref{fig:spectra_Euler} we present the kinetic energy spectrum for the same computational times shown in Fig. \ref{fig:vort_snaps}. 
As time evolves, energy is transferred to small scales where an approximate $l^{-3}$ scaling is established. The $-3$ exponent is consistent with the theory of Kraichnan \cite{Kraichnan1967} on the direct cascade of enstrophy of two-dimensional turbulence, extensively investigated numerically (see Bofetta~\cite{Boffetta2010} and references therein) and recently also revisited in the work of Cifani et.~al.~\cite{Cifani2022}. 
Moreover, recent studies by Modin and Viviani~\cite{Modin2021canonical} suggest that, for even longer times, the $-3$ exponent is replaced by the $l^{-1}$ scaling at small scales. 
It is outside the scope of this paper to investigate thoroughly such phenomenon, which we leave for future studies. 
Here, we show that the developed numerical method and parallel algorithms allow for the simulation of Lie--Poisson flows at unprecedented high resolutions while conserving the underlying Poisson structure. 

\begin{figure}[hbt!]
\centering
\includegraphics[width=0.6\textwidth]{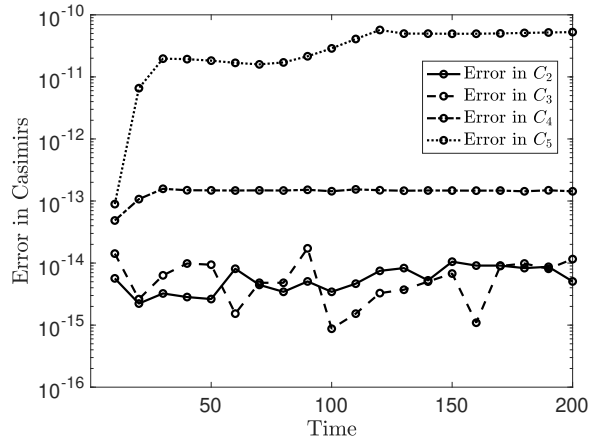}
\caption{Relative error of the Casimir functions $C_k$ for $k=2,...,5$ as a function of time. 
The tolerance for the fixed point iteration (\ref{eq:isomp_iter}) is set to $10^{-12}$.}
\label{fig:err_cas}
\end{figure}

The relative error of the Casimir functions $C_k$ (\ref{eq:Casimirs}) with respect to their values at the initial time for $k=2,...,5$ is shown in Fig.~\ref{fig:err_cas}.
For this simulation the tolerance for the fixed point iteration (\ref{eq:isomp_iter}) has been set to $10^{-12}$. 
For all Casimirs considered the error does not exceed $10^{-10}$ at the final time. 
For some $C_k$ a sub-linear accumulation of error is observed. Similar trends were found for higher order $C_k$.

\section{Conclusions}  \label{sec:con}
The main achievement documented in this paper is the development of a computationally efficient geometric integrator for Lie--Poisson flows on the sphere. 
We have extended the isospectral integrator \cite{Viviani2020} to high performance computing and demonstrated its capabilities for the simulation of high-resolution two-dimensional ideal flows on the sphere. 
This was accomplished by exploiting a tridiagonal splitting of the discrete spherical Laplacian in combination with highly efficient and scalable numerical algorithms. 
In particular, we overcome the problem of computing the discrete basis for the Lie algebra from its explicit expression in terms of the Wigner 3j-symbols \cite{Hoppe1989,Zeitlin2004}, which involves large factorials that are problematic to handle computationally. 
We reformulated, instead, the construction of the quantised basis as $\mathcal{O}(N)$ tridiagonal eigenvalue problems of size $\mathcal{O}(N)$. 
This allowed us to simulate for $N$ as large as $4096$. 
The same tridiagonal splitting was then used for the computation of the stream matrix. 
As a result, we have effectively reduced the complexity of the latter from the inversion of an $N^2 \times N^2$ matrix to that of the inversion of a set of tridiagonal matrices with overall complexity $\mathcal{O}(N^2)$, thus scaling optimally with respect to the $N^2$ spatial degrees of freedom.  

Particular attention has been given to the parallel implementation of $\Delta_N^{-1}$. Subsets of tridiagonal systems are assigned to each MPI process and solved locally by the optimised linear algebra package LAPACK. Data communication required to assemble such systems was carried out by constructing indexed derived MPI data types. The latter were employed in order to map the memory layout of the diagonals of the vorticity matrix and used in combination with non-blocking communication instructions. As a result, data transfer is completed with the minimum amount of MPI calls avoiding buffering. We found this to be essential for parallel scalability. 

Having reduced the computation of the inverse Laplacian to $\mathcal{O}(N^2)$ operations, the overall complexity of the method is that of matrix multiplication in the commutator, which is $\mathcal{O}(N^3)$. 
Dense matrix multiplications were carried out by employing the linear algebra library for distributed memory ScaLAPACK. The latter operates on a block-cycling decomposition, which optimises load-balance and communication across processors. The scaling tests, which we performed on two matrix sizes, $N=2048$ and $N=4096$, showed approximately linear scaling up to around $2500$ cores. Combining MPI with multi-threading appears to be advantageous, attaining the lowest computational time per time-step of $0.55$ seconds for $N=4096$. 
This computational time allows for long-time simulations and gathering of statistical quantities of complex flows.

As an illustration of the capabilities of the developed algorithm, we simulated Euler's equations for a matrix size of $2048\times 2048$. The simulation was carried out for $10^6$ time-steps corresponding to a total of $200$ time units. Conservation of the Casimir functions is shown to be robust over the entire simulation time, increasing only sub-linearly for some of the Casimirs.

In summary, we have developed the first Lie--Poisson integrator able to simulate fluid flows that span over a wide range of scales of motion. 
We expect this tool to be valuable in long-time studies of two-dimensional ideal fluid dynamics, particularly for a better understanding of how conservation of the Lie--Poisson structure and the Casimirs affects the qualitative behaviour. 
% Ultimatelly, this bridges the gap between numerical simulation of fluids and geometric integration. 

\section*{Acknowledgements}
This publication is part of the project SPRESTO which is financed by the Dutch Research Council (NWO). 
This work was also supported by the Swedish Research Council, grant number 2017-05040, and the Knut and Alice Wallenberg Foundation, grant numbers WAF2019.0201 and KAW2020.0287.
Simulations were carried out on the Dutch national e-infrastructure with the support of SURF Cooperative. 
We thank Francesco Viola for providing access to the cluster GALILEO100.

\bibliographystyle{unsrt}
\bibliography{mybibfile}

\end{document}